\documentclass[superscriptaddress,reprint,amsmath,amssymb,floatfix,aps]{revtex4-2}
\usepackage{graphicx}
\usepackage{enumitem}
\usepackage{xcolor}
\usepackage[normalem]{ulem}
\usepackage[fontsize=10pt]{scrextend}
\usepackage{graphicx}
\usepackage{wrapfig}
\usepackage{framed}
\usepackage{color}
\usepackage{float}
\usepackage{multirow}
\usepackage{xcolor}
\usepackage{soul}
\usepackage{amsmath} 
\usepackage{soul} 
\makeatletter

\newcommand*{\rom}[1]{\expandafter\@slowromancap\romannumeral #1@}
\makeatother

\usepackage{graphicx}
\usepackage{dcolumn}
\usepackage{bm}
\usepackage[version=4]{mhchem} 

\begin{document}
\preprint{APS/123-QED}
\title{Motility and interfacial instability of confined chemically active droplets }

\author{Pawan Kumar}
\affiliation{Department of Chemical Engineering, Indian Institute of Technology Kanpur, Kanpur, India}
\affiliation{Department of Basic Science and Humanities, Maharana Pratap Engineering College, Kanpur, Uttar Pradesh, India}

\author{Sobiya Ashraf}
\affiliation{Department of Chemical Engineering, Indian Institute of Technology Kanpur, Kanpur, India}

\author{Naveen Tiwari}
\affiliation{Department of Chemical Engineering, Indian Institute of Technology Kanpur, Kanpur, India}

\author{Dipin Pillai}
\affiliation{Department of Chemical Engineering, Indian Institute of Technology Kanpur, Kanpur, India}

\author{Rahul Mangal}
\email{mangalr@iitk.ac.in}
\affiliation{Department of Chemical Engineering, Indian Institute of Technology Kanpur, Kanpur, India}

\begin{abstract}
Microorganisms navigating through narrow spaces encounter significant hydrodynamic challenges. To overcome these constraints and sustain efficient motion, they employ adaptive strategies, including adaptive oscillatory body deformations. While artificial microdroplets can traverse channels narrower than their diameter, studies of their locomotion have thus far been largely restricted to steady-shape regimes. In this work, we demonstrate a transition from steady shape to dynamic interfacial undulations in 5CB (4'-Pentyl-4-cyanobiphenyl) droplets within aqueous trimethylammonium bromide (TTAB) solutions. We show that while droplets in dilute, additive-free solutions maintain a steady shape, the introduction of solutes or higher surfactant concentrations triggers pronounced interfacial undulations. Notably, both steady and undulating droplets exhibit a comparable velocity dependence on the confinement ratio—characterized by an initial deceleration followed by saturation—governed by the competition between hydrodynamic resistance and phoretic flow within the lubrication film. Furthermore, we find that increased surfactant concentration increases the Capillary number, resulting in a thicker lubrication layer that facilitates a symmetry-breaking transition. Upon varying confinement, the droplet interface shifts from bilateral undulations to being localized on one side, traveling-wave mode strongly coupled to flow field fluctuations at the droplet’s anterior. Linear stability analysis identifies the Yih–Marangoni instability as the underlying mechanism for these oscillations, revealing a previously unrecognized mode of adaptive locomotion in confined active matter.

\end{abstract}

\maketitle

\section{Introduction}

The locomotion of microorganisms is strongly influenced by the geometric constraints of their environment. In natural and physiological settings—ranging from soil pores and aquatic microchannels to vascular and intracellular spaces—self-propelled organisms such as bacteria, euglenas, and protozoa frequently navigate confined geometries that regulate their motility and, in turn, impact processes such as nutrient acquisition and survival \cite{jin2024microbes, yang2018mechanisms, fenchel2008oxygen}. Notably, these swimmers often avoid direct contact with confining boundaries and are instead separated by a thin lubricating fluid film that influences local hydrodynamics, provides mechanical cushioning, and facilitates navigation. Under extreme confinement, microorganisms lacking effective flagellar propulsion may adopt alternative survival strategies, including elongation, division, or significant deceleration. For example, Euglena gracilis transitions from flagellar swimming to periodic shape oscillations driven by sliding pellicular strips when navigating tapered capillaries. These cyclic deformations generate localized bulges that press against the capillary walls, enabling an efficient crawling mode of locomotion \cite{noselli2019swimming}. Similarly, protozoa may elongate to preserve mobility \cite{wang2005mobility}, while bacteria can exploit wall-induced hydrodynamic interactions under moderate confinement to enhance their swimming speeds \cite{wioland2016directed, wei2025confinement, lynch2022transitioning}. 

Artificially swimming liquid droplets have emerged as robust model systems for mimicking such biological behaviors. Liquid crystal (LC) droplets, in particular, provide a versatile platform owing to their intrinsic anisotropy, phase-transition, and responsiveness to external stimuli. Their autonomous motion is typically driven by interfacial Marangoni gradients arising from micellar solubilization in surfactant-rich environments \cite{dwivedi2021solute, izzet2020tunable, izri2014self}. Briefly, empty micelles solubilize the 5CB droplet, releasing oil in the form of dispersed nanometer-sized filled micelles into the surrounding vicinity. This process induces a spatial asymmetry in the surfactant monomer concentration at the droplet–solution interface. The resulting gradient generates Marangoni stresses, driving an interfacial flow from regions of low interfacial tension (associated with high surfactant monomer concentration) toward regions of high interfacial tension (low surfactant monomer concentration). Consequently, the droplet undergoes net forward propulsion toward regions rich in empty micelles, while a trail of filled micelles—hereafter referred to as solute—is left behind to satisfy the conservation of linear momentum. The nonlinear coupling between the advective and diffusive time scales of the solute field, captured by non-dimensional Péclet number $Pe$, around the droplet plays a crucial role in governing the swimming mode \cite{morozov2019nonlinear}. Over the past decade, extensive research has uncovered the rich diversity of self-propulsion behaviors exhibited by active droplets. Depending on system parameters, particularly $Pe$, isolated droplets transition from straight to curvilinear to jittery trajectories \cite{suda2021straight, dwivedi2021solute, hokmabad2021emergence, dwivedi2023mode}, or can undergo nematic elasticity–driven curling behavior \cite{kruger2016curling, suga2018self}. More complex responses, such as auto-negative chemotaxis \cite{jin2017chemotaxis}, upstream rheotaxis in external flows \cite{dwivedi2021rheotaxis, dey2022oscillatory}, electrotaxis \cite{buness2024electrotaxis}, magnetotaxis \cite{wagner2024magnetotaxis},  temperature switchable activity \cite{kumar2025temperature, ramesh2025frozen}, and deformation in viscoelastic media \cite{dwivedi2023deforming} further highlight their adaptive nature. Recent efforts have shifted toward understanding pair-wise to collective dynamics, where droplet–droplet interactions give rise to emergent behaviors such as scattering \cite{dwivedi2025chemical}, chasing \cite{kumar2024motility}, predator–prey motion \cite{meredith2020predator}, disorder to order transition \cite{ashraf2025emergence}, rotating clusters \cite{hokmabad2022spontaneously}, dynamic assemblies \cite{thutupalli2018flow}, and chemotactic self-caging \cite{hokmabad2022chemotactic}. Collectively, these findings emphasize how chemical activity and hydrodynamic coupling orchestrate a broad spectrum of nonlinear phenomena in active droplet systems.

The dynamics of active droplets are strongly influenced by geometric confinement, which modifies both interfacial activity and the surrounding flow fields across different confinement regimes. Under strong quasi-2D confinement, increasing droplet size (and hence $Pe$) in disk-shaped (S)-4-cyano-4’-(2-methylbutyl)biphenyl (CB15) droplets drives transitions from dipolar to quadrupolar and higher-order modes, accompanied by a systematic posterior-to-anterior migration of vortices \cite{ramesh2023interfacial}. This behavior highlights the role of slow micellar diffusion and chemical buildup in governing hydrodynamic modes, as well as the long-time arrest of activity due to interfacial saturation by reaction products. In one-dimensional setting, where the channel width exceeds the droplet size, recent studies have shown that both hydrodynamic and chemical fields are modulated by $Pe$ and the degree of confinement, leading to variations in swimming behavior, such as motion along channel walls or the channel centerline \cite{kumar2024motility, singh2025active}. However, relatively few studies have addressed regimes in which the droplet size exceeds the capillary width, offering strong 1-D confinement. In a seminal work, \citet{de2021swimming} demonstrated that self-propelled water-in-oil droplets within capillaries can sustain motion even under strong confinement, with the swimming velocity saturating once the droplet length exceeds the capillary height. Under such conditions, the lubrication film surrounding the droplet becomes non-uniform, giving rise to non-axisymmetric flow circulation. In tapered capillaries, this asymmetry can induce the formation of a rear neck that deepens and ultimately leads to spontaneous droplet division. More recently, Guchhait \textit{et al.} reported non-axisymmetric hydrodynamic signatures in the motion of liquid-crystal droplets under confinement \cite{guchhait2025flow}. With increasing confinement, droplets deform from spherical to stadium-like and eventually to elongated capsule-like shapes, accompanied by a transition in flow fields from pusher-type to puller-type and ultimately to strongly asymmetric velocity distributions. Despite these advances, studies of confined active droplets remain limited and have primarily focused on steady interfacial configurations characterized by thin lubrication films. Investigating such behaviors in artificial microswimmers under confinement not only provides insight into microbial locomotion in complex environments but also informs applications in targeted drug delivery and lab-on-a-chip technologies. \\

In this work, we demonstrate for the first time, to the best of our knowledge, that confined LC droplets swimming in capillaries can exhibit interfacial traveling waves, revealing a previously unexplored mode of active droplet locomotion under confinement. The results of our work are organized into two sections to systematically explore the dynamics of confined active droplets. In the first part (section IIIA), we present the dynamics of droplet under varying confinement ratios and physicochemical environments, emphasizing how the addition of solute to surfactant influences the swimming velocity and spatial distributions of the flow and chemical fields surrounding the confined droplets. In the second part (section IIIB), we discuss the emergence of interfacial traveling waves under distinct solute and surfactant concentrations, revealing a new regime of dynamic shape fluctuations exhibited by active droplets. In the third part (section IIIC), the underlying mechanism of instability is discussed.

\section{Materials and Methods}
\begin{figure*}[t]
 {\includegraphics[scale=0.7]{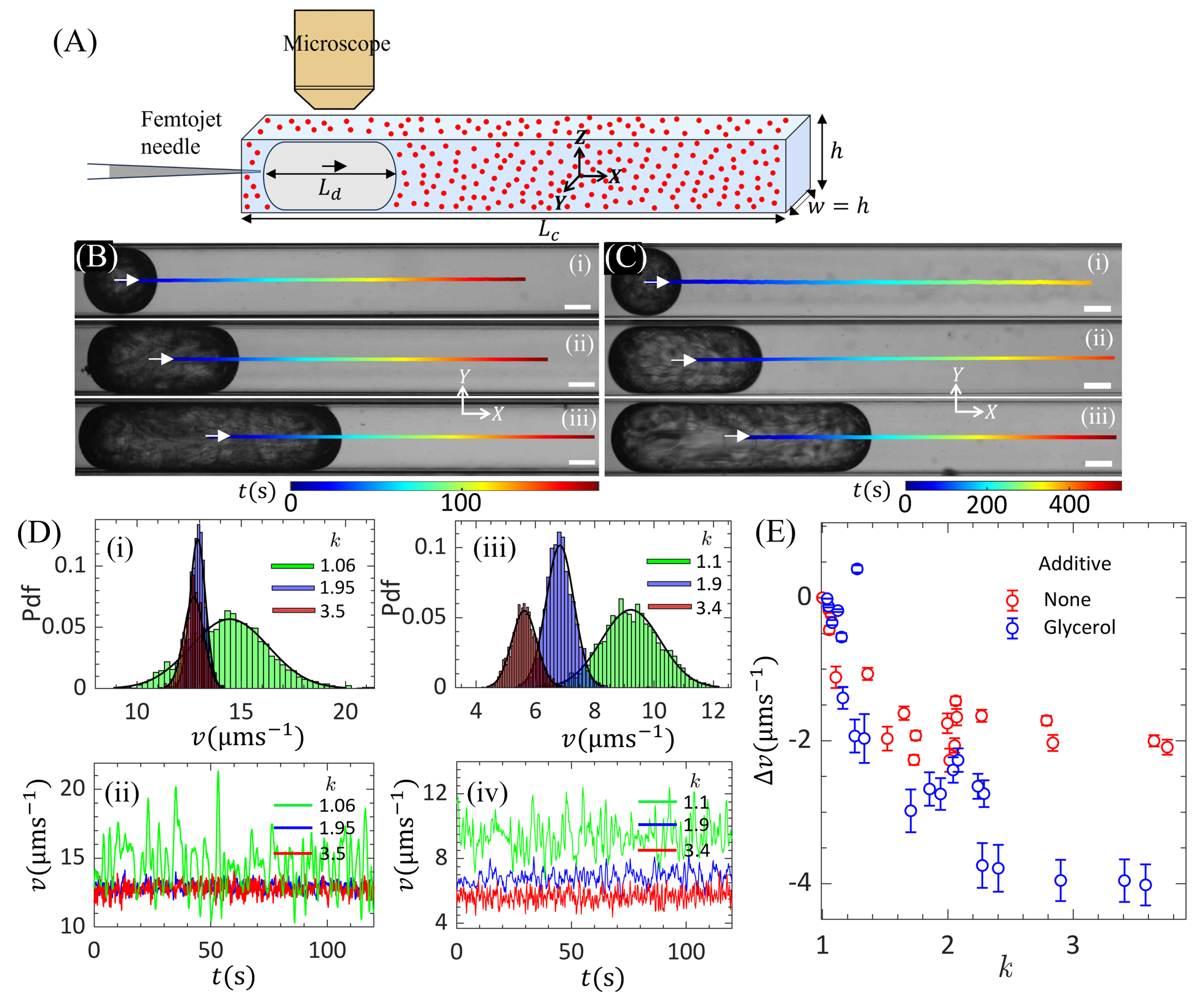}}
\caption{\small (A) Schematic of the square capillary channel used for 1-dimensional experiments. Representative $x-y$ trajectories superimposed on optical micrograph of a 5CB swimming droplet in 6 wt.$\%$ aqueous TTAB solution (B) without additive and (C) with 80 wt.$\%$ glycerol as additive for three different confinement ratios $k$, (i) 1.03, (ii) 2.0 and (iii) 3.6. Scale bars' length is 200 $\mu$m. (D) Probability density distributions and time-series of instantaneous velocities of 5CB droplet in 6 wt.$\%$ TTAB aqueous solution (i and ii) without additive and (iii and iv) containing 80 wt. $\%$ glycerol as an additive for three different confinement ratios. (E) Difference in swimming velocity of droplets with respect to the droplet velocity at $k$ $\sim$ 1 as a function of confinement ratio.}
\label{fig1}
\end{figure*}

As shown in figure~\ref{fig1}(A), a square glass capillary with edge length $h = 500~\mu$m and axial length $L \sim 3$–$5$ cm was used as the optical chamber. The capillary was filled with an aqueous solution of 6 wt.$\%$ trimethylammonium bromide (TTAB; Loba Chemicals), which served as the surfactant medium. Using a micro-injector (FemtoJet 4i, Eppendorf), oil droplets of 4-cyano-4$^\prime$-pentylbiphenyl (5CB; Jiangsu Hecheng Advanced Materials Co., Ltd.) with varying diameters $>$ 500~$\mu$m were introduced into the surfactant-filled capillary. As the droplet diameter exceeds the channel width, the droplets undergo lateral squeezing as shown in figures~\ref{fig1}(B) and \ref{fig1}(C). Increasing droplet size drives a morphological transition from stadium-like to capsule-like droplet shapes. Accordingly, we characterize this confined state by using confinement ratio as $k = L_d / w$, where $L_d$ represents the axial length of the confined droplet and $w$ denotes the channel width. Following droplet injection, both ends of the capillary were sealed with grease to suppress background convection. The self-propelled motion of the droplets was recorded in bright-field mode using a BFS-U3-70S7CC camera at 3–10 frames~s$^{-1}$ and 2–10$\times$ magnification on an upright optical microscope (Olympus BX53). To maintain isothermal conditions, the capillaries were mounted on a temperature-controlled stage maintained at $25^{\circ}$C. In selected experiments, the swimming dynamics of the droplets were modified by adding solutes to the 6 wt.$\%$ TTAB aqueous solution, including 20-80 wt.$\%$ glycerol (Loba Chemicals), 20 wt.$\%$ polyvinyl pyrrolidone (PVP; $M_w \sim 40$~kDa, TCI Chemicals), 1 wt.$\%$ polyacrylamide (PAAm; $M_w \sim 6000$~kDa, Sigma-Aldrich), and 50 wt.$\%$  sucrose (Sigma-Aldrich), to systematically tune the interfacial activity and transport properties. The droplet trajectories in the $x-y$ plane were recorded at 20 frames s$^{-1}$ using a FLIR camera (ORX-10G-71S7C-C) and analyzed by tracking the droplet centroids with the Mtrack2 plugin in ImageJ.\\

To investigate the fluid flow around the swimming droplet, particle image velocimetry (PIV) was performed using fluorescent polystyrene tracer particles (diameter $\sim$ 3.2 $\mu$m) dispersed in the continuous phase. These experiments were conducted on an inverted microscope (Olympus IX73) equipped with a fluorescence illuminator (U-RFL-T) and a 560 nm laser to excite the tracers. The velocity vectors of the tracers were characterized using PIVlab - an open-source particle image velocimetry tool with GUI in MATLAB \cite{thielicke2021particle}, and streamlines were plotted in Techplot 360. To probe the chemical field of filled micelles around the droplets, an oil-soluble fluorescent dye (Nile Red, Sigma Aldrich) was added to the droplet. ORX-10G-71S7C-C, FLIR, camera connected with the microscope (4-6.4X magnification), was used to record the fluorescence videos (20 frames s$^{-1}$). The interfacial tension of 5CB droplet in aqueous solutions containing varying concentrations of TTAB and solute environments were adopted from the work of \citet{dwivedi2021solute}. 
Finally, the bulk phase viscosity ($\eta_b$) of the continuous phase was determined using an IKA ROTAVISC lo-vi S000 viscometer, while the 5CB droplet phase viscosity ($\eta_d$) was measured using an Anton Paar MCR 302e rheometer with 25 mm parallel plate geometry.

\section{Results and discussion}

\subsection{Dynamics of confined droplets}

\begin{figure*}[t]
{\includegraphics[scale=0.5]{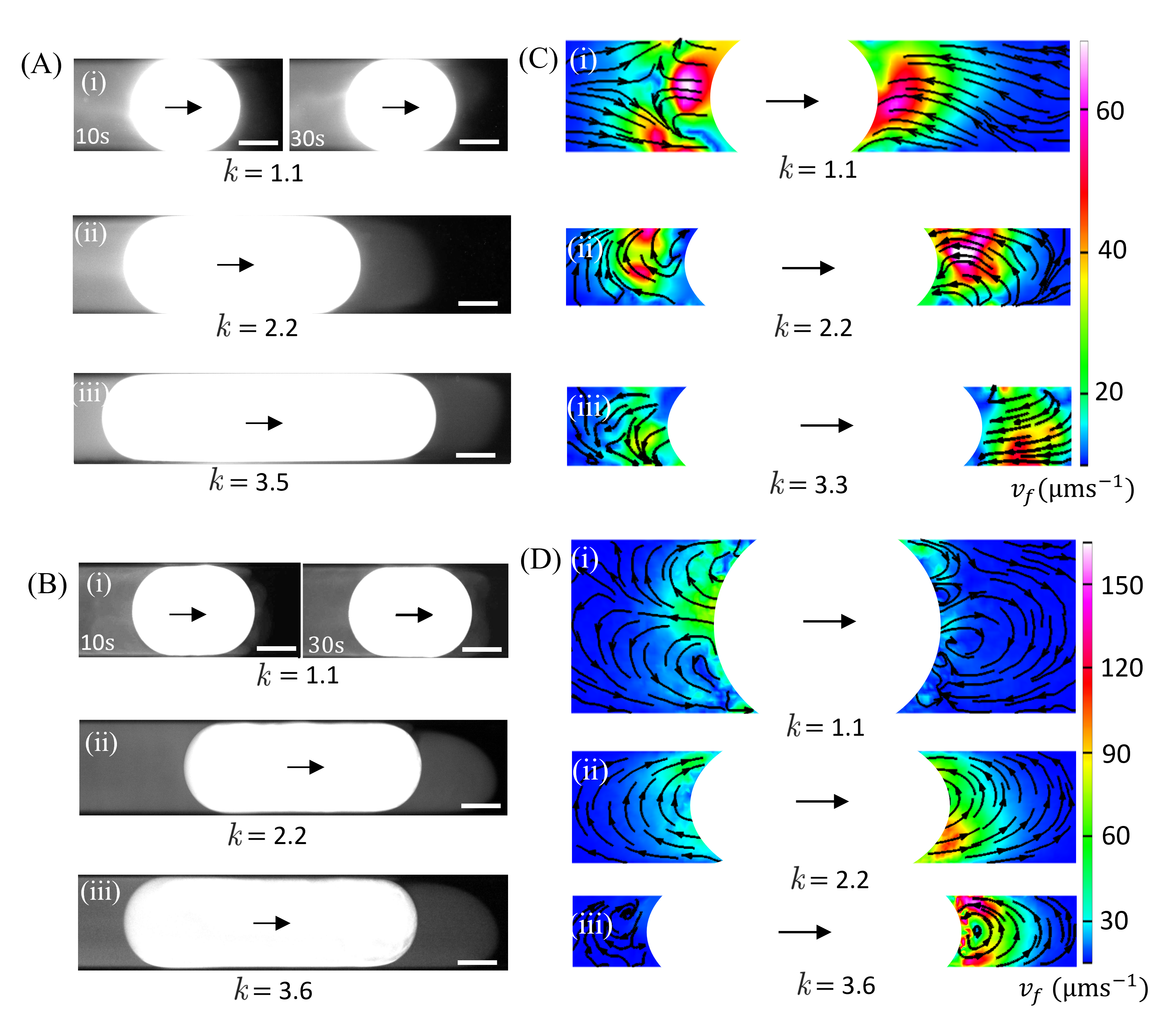}}
\linespread{1.0}
\caption{\small (A, B) Optical micrographs of chemical field distribution and (C, D) flow velocity distributions around the confined droplet for three distinct confinement ratios in 6 wt.$\%$ TTAB aqueous solution (A, C) without additive and (B, D) with 80 wt. $\%$ glycerol as an additive. The scale bars correspond to 200 $\mu$m.}
\label{fig2}
\end{figure*}

We first investigated the dynamics of confined 5CB droplets in two distinct continuous phases: a 6 wt.\% TTAB aqueous solution and a 6 wt.\% TTAB solution supplemented with 80 wt.\% glycerol, across a range of confinement ratios $k$.  In both systems, the surfactant concentration was maintained significantly above the critical micelle concentration (CMC $\approx$ 0.13 wt.\%), ensuring the immediate onset of self-propulsion via micelle-mediated solubilization upon injection. In all cases, the droplets migrated toward the distal end of the capillary. Representative optical micrographs, overlaid with the corresponding $x$–$y$ trajectories of the droplet centroids (Figures~\ref{fig1}(B, C)), demonstrate that the droplets undergo predominantly rectilinear motion along the $+x$ direction regardless of the confinement ratio. This unidirectional motion is further evidenced by the unimodal velocity distributions (Figures~\ref{fig1}(D(i), D(iii))), which are strictly restricted to positive values. Under weak confinement ($k \approx 1$), the droplets exhibit pronounced velocity fluctuations (Figures~\ref{fig1}(D(ii, iv))), resulting in a relatively broad distribution of instantaneous speeds. However, these fluctuations are progressively suppressed as the degree of confinement increases. Specifically, stronger confinement restricts transverse ($y$-direction) excursions, leading to narrower velocity histograms and a more stable, nearly constant propulsion speed in the high-confinement regime.

Figure~\ref{fig1}(E) illustrates the influence of the confinement ratio, $k$, on the relative swimming velocity ($\Delta v$) of confined 5CB droplets compared to the $k \approx 1$ baseline. In both systems, $\Delta v$ initially decreases with increasing confinement, reaching a plateau at $k \approx 1.5$ for the aqueous solution and $k \approx 2.5$ for the glycerol-supplemented medium. Beyond these values of $k$, $\Delta v$ plateaus at approximately $-2~\mu\mathrm{m\,s^{-1}}$ and $-4~\mu\mathrm{m\,s^{-1}}$, respectively. For droplets under strong confinement ($k > 1$), a thin lubrication film ($\sim$100nm - $10 ~\mu\mathrm{m}$) develops between the droplet interface and the capillary walls. While this film prevents direct wall contact, it substantially increases hydrodynamic resistance, thereby reducing the droplet velocity. Furthermore, confinement significantly enhances the accumulation of the chemical field at the anterior (frontal) side, resulting in a persistent chemical plume (see optical micrographs in figures~\ref{fig2}(A, B). While this localized chemical buildup promotes deceleration, the saturation of $\Delta v$ observed at higher confinement suggests the presence of a concurrent accelerating mechanism. Theoretical framework of \citet{guchhait2025flow}, demonstrates that increasing $k$ enhances the interfacial tension at the posterior edge relative to the anterior. This intensified Marangoni stress gradient increases the interfacial slip velocity, $u_s$, from the anterior toward the posterior—an effect that, in isolation, would drive an increase in propulsion speed. Consequently, the observed velocity plateau reflects a dynamic equilibrium between escalating viscous resistance and enhanced Marangoni stress-driven flow. The saturation of the droplet velocity beyond a threshold confinement ratio thus marks the point where these competing hydrodynamic and phoretic influences reach a steady balance.\\

Despite qualitative similarities in variation of $\Delta v$ with confinement in both cases, the quantivative differences, both in terms of $\Delta v$ and transition values of $k$, is attributed to an approximately 40-fold increase in the Péclet number ($Pe$) upon the addition of 80 wt.$\%$ glycerol to the 6 wt.$\%$ TTAB solution which indicates a strong enhancement of advective transport relative to diffusion. This difference in underlying $Pe$ is more apprarent in the flow-fields generated by the droplets. In aqueous TTAB solution, the lower $Pe$ regime results in a characteristic puller-like swimming mode (figure~\ref{fig2}(C(i))). Conversely, in the 80 wt.\% glycerol solution, the enhanced advective transport leads to a flow field in the anterior region characterized by multiple small, time-fluctuating circulation zones (see figure~\ref{fig2}(D(i)) and figure S1 in supplementary information), resembling a symmetric quadrupolar flow structure. As the confinement is increased, the flow field in both systems, as indicated in figures~\ref{fig2}(C(ii, iii)) and (D(ii, iii)), becomes non-axisymmetric, giving rise to single, well-defined circulations in both the anterior and posterior regions of the droplet that remain stable over time. Such a non-axisymmetric distribution of the flow velocity has been reported in previous experimental and theoretical studies and has been suggested to arise from the nonuniform formation of the lubrication layer ~\cite{de2021swimming, guchhait2025flow}. In our case, the addition of 80 wt.$\%$ glycerol to the 6 wt.$\%$ TTAB solution leads to the emergence of a significantly asymmetric and undulatory lubrication layer (Section IIIB), which is expected to enhance the vorticity of the surrounding fluid circulation by nearly an order of magnitude compared to the 6 wt.$\%$ TTAB aqueous solution (see figure S2 in supplementary information).

\subsection{Emergence of Interfacial traveling waves in confined active droplets}
\begin{figure*}[t]
\centering {\includegraphics[scale=0.45]{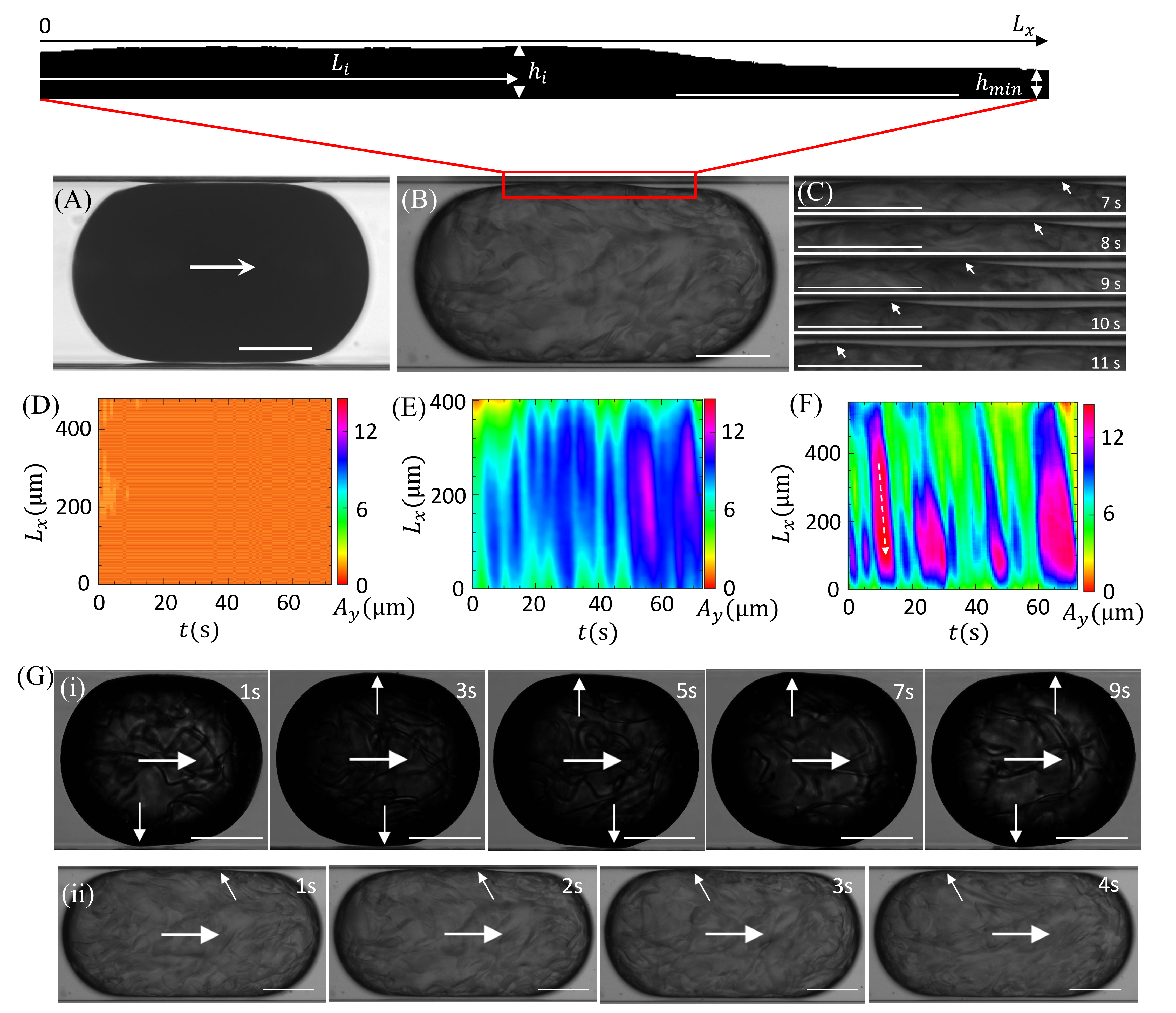}}
\caption{\small Optical micrograph of droplet swimming in 6 wt.$\%$ TTAB aqueous solution (A) without additive ($k=1.6$) and (B) with 80 wt.$\%$ glycerol ($k=1.9$). (C) Image sequence of the lateral cross-section (as shown in image 5B) of droplet swimming in 6 wt.$\%$ TTAB aqueous solution with 80 wt.$\%$ glycerol additive ($k=1.9$). Spatiotemporal plot of the deformations (D) without additive and (E, F) with 80 wt.$\%$ glycerol for the confinement ratio (E) $k=1.1$ and (F) $k=1.9$, and (G) shows the corresponding image sequence with time. Arrows near the droplet interfaces indicate the deformations of the droplet. The scale bars denote 200 $\mu$m.}
\label{fig3}
\end{figure*}

In 6~wt.\% TTAB aqueous solution, the droplet (see optical micrograph in Figure~\ref{fig3}(A)) propels while maintaining a steady morphology across all investigated values of $k$, with no discernible surface deformation observed. However, upon the introduction of 80~wt.\% glycerol into the solution, for all values of confinement ratio ($k > 1$), lateral interfacial oscillations emerge, as illustrated in figure~\ref{fig3}(B). These oscillations manifest as waves propagating from the anterior to the posterior of the droplet, a progression confirmed by the directional arrows in the time-sequence micrographs in figure~\ref{fig3}(C) and further demonstrated in Supporting Video S1. To quantify this interfacial dynamics, the oscillation amplitude $A_y$ is defined as the difference between the local interface height $h_i$ and the minimum height $h_{\text{min}}$ measured relative to a reference line along the cross-sectional length $L_x$, extending from the rear to the front of the droplet, excluding the curvature-dominated regions at the poles, as shown in the magnified view in Figure~\ref{fig3}(B). In the absence of glycerol, as expected the droplet exhibits uniform intensity of $A_y$ in the spatiotemporal plot (figure~\ref{fig3}(D)) indicating a stable morphology devoid of surface undulations. In contrast, the introduction of glycerol triggers pronounced traveling waves, captured in the spatiotemporal representations of $A_y$ for $k$ values of 1.1 and 1.9 in figures~\ref{fig3}(E) and (F), respectively, suggesting sensitivity to the degree of confinement. At lower confinement levels, oscillations appear to develop on both lateral interfaces of the droplet (see time lapse micrographs shown in figure~\ref{fig3}(E(i))). However, as the confinement ratio increases, these oscillations typically undergo symmetry breaking and localize to a single side of the droplet (see time lapse micrographs shown in figure~\ref{fig3}(E(ii))). 

This transition in the nature of the interfacial dynamics aligns with the emergent asymmetry observed in the associated chemical and flow fields (see Figure~\ref{fig2}(B, D)), suggesting that the mechanism driving these oscillations is intrinsically coupled to the surrounding chemical and flow fields. For elongated droplets confined within a square channel satisfying the condition $L Ca^{1/3} \ll 1$ (where $L$ is the dimensionless length of the droplet and $Ca = \eta_s U_{\text{rel}}/\sigma$ is the capillary number, which characterizes the relative importance of viscous shear to interfacial tension, with $\eta_s$, $U_{\text{rel}}$, and $\sigma$ denoting the continuous phase viscosity, the droplet's relative velocity, and the interfacial tension, respectively), the corner gutters serve as effective bypass channels for the continuous phase \cite{rao2018motion}. In our experiments, this condition is satisfied for both aqueous 6wt.\% TTAB case ($Ca \sim 10^{-6}$) and with 80~wt.\% glycerol ($Ca \sim 10^{-4}$). In the case of glycerol at lower confinement ratios, these corner gaps—characterized by lower hydraulic resistance—permit the flow field at the anterior side of the droplet to bypass the body, thereby facilitating the dispersion of the chemical field. However, at higher confinement ratios ($k$), the increased droplet length and associated shear drag significantly enhance the resistance to this bypass flow. This restricted passage leads to a localized accumulation of the chemical field, eventually triggering a symmetry-breaking event. Consequently, the droplet loses its axial symmetry, forcing the bypass flow to predominantly localize to a single side, which correlates with the observed localization of the interfacial oscillations.

\begin{figure*}[ht]
 \centering {\includegraphics[scale=0.4]{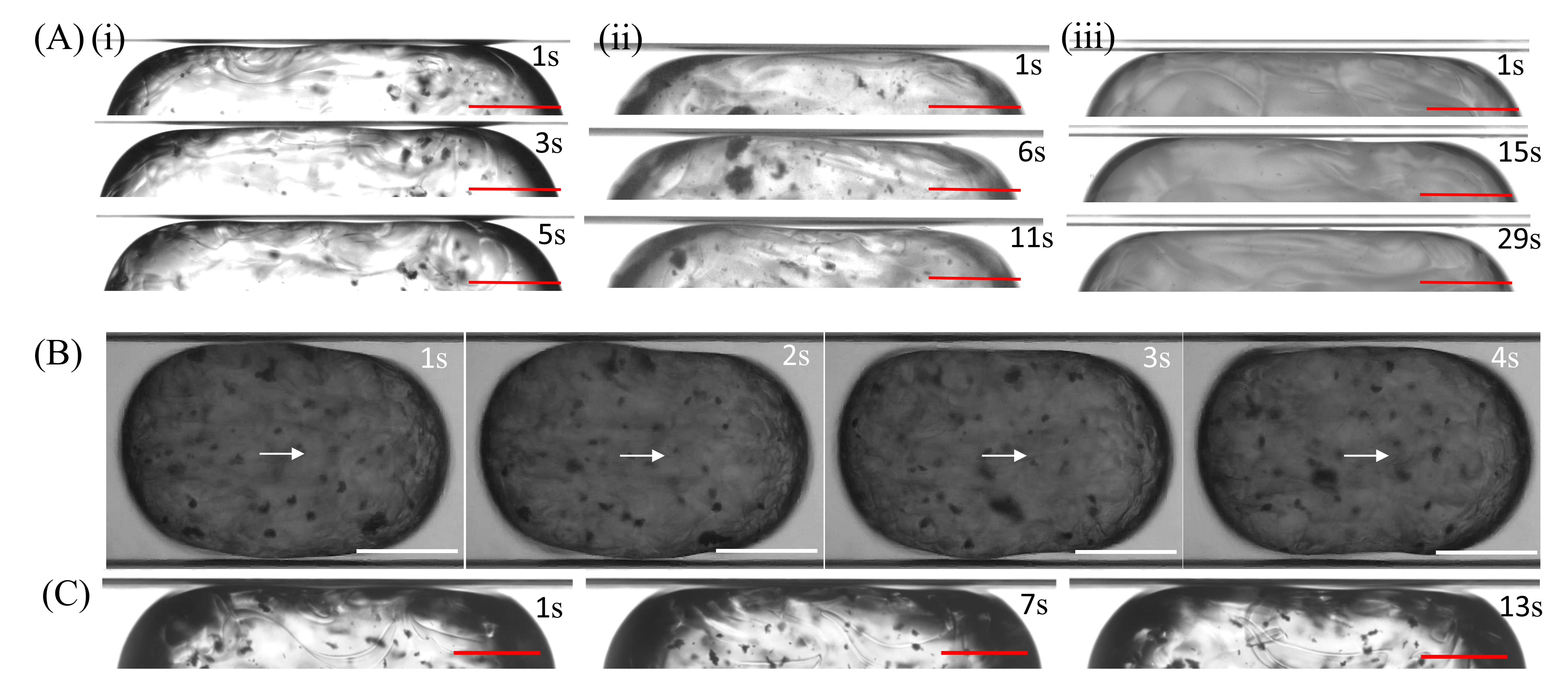}}
\caption{\small Optical micrograph of oscillating 5CB droplet image sequence with time (A) in 6 wt.$\%$ TTAB solution with additives (i) 20 wt.$\%$ PVP, (ii) 1 wt.$\%$ PAAm, (iii) 50 wt.$\%$ sucrose, (B) in 35 wt.$\%$ TTAB aqueous solution and (C) in 6 wt.$\%$ TTAB aqueous solution containing 0.5 wt.$\%$ PEO for $k = 1.6-1.9$. All scale bars are 200 $\mu$m.}
\label{fig4}
\end{figure*}

Next, to verify the role of $Pe$ in the observed interfacial oscillations, we investigated the droplet dynamics by adding various solutes—20~wt.\% of 40kDa Polyvinylpyrrolidone (PVP), 1~wt.of 6000kDa \% Polyacrylamide (PAAm), 0.5 wt.\% 8000 kDa PEO (Polyethylene Oxide)—to a 6~wt.\% TTAB aqueous surfactant solution, as well as in a 35~wt.\% TTAB solution (no additive) subjected to a confinement ratio of $k \sim 1.6-1.9$. As illustrated by the chronological micrographs in figure \ref{fig4}(A) and (B) droplets in these systems exhibit traveling waves along their interfaces (see Supporting Videos S2-S5). In the 35~wt.\% TTAB solution, oscillations are observed on both sides of the droplet across all investigated confinement ratios (see representative time-lapse micrographs shown in figure~\ref{fig4}(B) for $k=1.6$). This behavior is consistent with the observed surrounding fluid flow at both lateral interfaces with temporal fluctuations (see figure S3 in supplementary information). Compared to the 80~wt.\% glycerol case, the absence of chemical field asymmetry—and the resulting lack of flow-field asymmetry—is consistent with the increased gap thickness observed in the 35~wt.\% TTAB solution. Specifically, the mean gap thickness $\langle e \rangle \sim 22~\mu\text{m}$ in the latter case is significantly greater than the $\sim 10~\mu\text{m}$ range observed in the former. This increased gap thickness is consistent with the higher $Ca$ values ($\sim$ 10$^{-4}$ for the 80~wt.\% glycerol case vs. $\sim$ 10$^{-3}$ for the 35~wt.\% TTAB case). The underlying scaling laws and the resulting hydrodynamic implications will be discussed in detail in the subsequent section. Furthermore, droplets in 6~wt.$\%$ TTAB aqueous solution containing 0.5~wt.\% 8000~kDa PEO as an additive is observed to exhibit interfacial waves, as shown in figure~\ref{fig4}(C), at a significantly lower Péclet number (see Table~\ref{table1}). This finding suggests that the onset of interfacial oscillations is not a simple consequence of high $Pe$; rather, the phenomenon appears to be governed by alternative flow-related physics, which we discuss in the following section.

\begin{table}[th]
   \caption{$Pe$ of active 5CB droplet in different solutions for the confinement ratio $k$ = 1.8$\pm$0.2. The effective size $a$ of the confined droplet is given by $\sqrt{L_d{_x}L_d{_y}}$, where $L_d{_x}$ and $L_d{_y}$ are the lengths of the droplet along the major and minor axes, respectively.}
   \setlength{\tabcolsep}{0.7\tabcolsep}
    \centering
  \begin{tabular}{c c c c c c c c}
    \hline
    c$_\text{TTAB}$ & Additive & $a$ & $\langle v \rangle$ & {$D$} & $Pe$ & {Mode}  \\
    
    wt.$\%$ & Solute & $\mu$m & $\mu$m s$^{-1}$& $\mu$m$^{2}$ s$^{-1}$& & & \\
    
    \hline
    6 & PVP (20 wt.$\%$)  & 670 & 8 & 1.02 & 5245 & Yes \\
    6 & PAAm (1 wt.$\%$)  & 670& 4.6 & 0.91 & 3375 &Yes\\
    6 & Sucrose (50 wt.$\%$)&670& 2.5 & 1.4 & 1163 & Yes\\
     6 & PEO (0.5 wt.$\%$)  &670& 3.7 & 25 & 95 & Yes\\
      6 & None  &670& 13.5 & 25 & 350 & No\\
       \hline
    \end{tabular}
    \label{table1}
     \end{table}

For motion of an elongated droplet through a viscous medium in a microcapillary channel is closely related to the classical Bretherton problem, in which the thin lubrication film separating the droplet from the capillary wall plays a critical role in determining both the pressure distribution and the migration velocity. As discussed earlier using lubrication theory, classical Bretherton scaling that $e/h$ $\sim$ $Ca^{2/3}$ \cite{bretherton1961motion}. Subsequent studies have demonstrated that the presence of soluble or insoluble surfactants generates interfacial concentration gradients, giving rise to Marangoni stresses that substantially modify the lubrication dynamics. These stresses enhance the film thickness \cite{olgac2013effects}, as surfactant depletion within the thin-film region induces Marangoni-driven flow into the gap. For infinitely long bubbles, this mechanism leads to an increase in film thickness by a factor of $4^{2/3}$ compared to surfactant-free cases \cite{ratulowski1990marangoni}. Utilizing the observed dynamics of confined active droplets across the previously investigated systems, alongside the aqueous TTAB solutions (with $c_{\text{TTAB}}$ ranging from 6-35~wt.\%), we evaluated the key dimensionless physical parameters. Specifically, we focused on the viscosity ratio $m = \eta_{\text{LC}}/\eta_s$ (where $\eta_{\text{LC}}$ and $\eta_s$ denote the dynamic viscosities of the liquid-crystal droplet and the continuous phase, respectively) and the capillary number $Ca$. Figure~\ref{fig6}(A) shows that the normalized film thickness $\langle e \rangle/h$ increases with $Ca$ and follows a scaling exponent $\alpha = 0.25 \pm 0.08$. 
The observed deviations in film-thickness growth compared to previous surfactant-based studies~\cite{ratulowski1990marangoni} are likely attributable to fundamental differences in the underlying assumptions. Specifically, lubrication-theory analyses typically assume droplets with steady interfacial configuration, whereas our measurements involve active droplets with dynamically evolving and unstable interfaces. Furthermore, these are force-free active swimmers, in contrast to the classic Bretherton scaling, which describes externally forced droplets or bubbles. Under regimes where the droplet interfaces remained stable, they maintained near-wall contact; due to the limited spatial resolution required to resolve such ultra-thin lubrication films, these cases were excluded from the film-thickness analysis. The relevant physical parameters for all investigated systems are summarized in Table \ref{table2}.

\begin{table*} 
   \caption{Characteristics of 5CB droplet in different solutions for the confinement ratio $k$ = 1.8$\pm$0.2.}
   \setlength{\tabcolsep}{0.7\tabcolsep}
    \centering
  \begin{tabular}{c c c c c c c c c c c c c c c c c c c c c c c c c c c c }
    \hline
    c$_\text{TTAB}$ & Additive & $v$ & $\langle v_f \rangle$ & {$ \text{$\langle U_{rel} \rangle$}$} & $\xi$ & $\eta{_s}$ & $m$ & {$\sigma$}\cite{dwivedi2021solute} & {Ca}& {Mode}  \\
    
    wt.$\%$ & Solute & $\mu$m s$^{-1}$ & $\mu$m s$^{-1}$& $\mu$m s$^{-1}$& $\mu$m& mPa.s & & mNm$^{-1}$& ($\times$10$^{-3}$)& \\
    
    \hline
    6 & None  & 13.5 & 52 & 66 & X& 1.66 &15.06 &30&0.0037 &No  \\
    10& None  & 17.3 & 62 & 79 & X& 2.02 &12.38 &30&0.0053 &No\\
    15& None  & 15.3 & 43 & 58 & X& 3.01 &8.31  &30&0.0058 &No \\
    20& None  & 15.9 & 43 & 69 & 7& 5.1  &4.9   &30&0.013 &Yes \\
    25& None  & 14.9 & 32 & 47 & 5& 12.6 &1.98  &30&0.02  &Yes\\
    30& None  & 12.3 & 29 & 41 & 10& 63.4 &0.39  &30&0.087 &Yes \\
    35& None  & 12.2 & 23 & 35 &  15& 380  &0.066 &30&1.33 &Yes \\
6 & Glycerol (80 wt.$\%$)& 7.2 & 71 & 78 &9 &  43 &0.58 &10 &0.34&Yes \\
6 & Glycerol (20 wt.$\%$)&18.3 & 45 & 63 & X & 2.48 &10.1 &10 &0.07&No\\
     3.5 & Glycerol (40 wt.$\%$)&13.9 & 31 & 45 & X & 4.26 &5.87 &10 &0.09&No\\
     3 & Glycerol (50 wt.$\%$)&13.6 & 32 & 46 & X & 5.9 &4.24 &10 &0.03 &No\\
     2.75 & Glycerol (55 wt.$\%$)&12.2 & 23 & 35 & X & 8.2 &3.05 &10 &0.03&No\\
     3 & Glycerol (60 wt.$\%$)&12.5 & 35 & 48 & 5 & 11.1 &2.25 &10&0.05&Yes\\
    6 & PVP (20 wt.$\%$)  & 8 & 43 & 51 & 7.7 & 42 &0.6 &20 &0.1&Yes \\
    6 & PAAm (1 wt.$\%$)  & 4.6& 31 & 35.6 & 8.2 & 47 &0.53 &20 &0.08&Yes\\
    6 & Sucrose (50 wt.$\%$)&2.5& 11 & 13.5 & 8.2 & 29.8 &0.8 &-- & -- &Yes\\
          \hline
    \end{tabular}
    \label{table2}
     \end{table*}

\begin{figure}[ht]
 \centering {\includegraphics[scale=0.7]{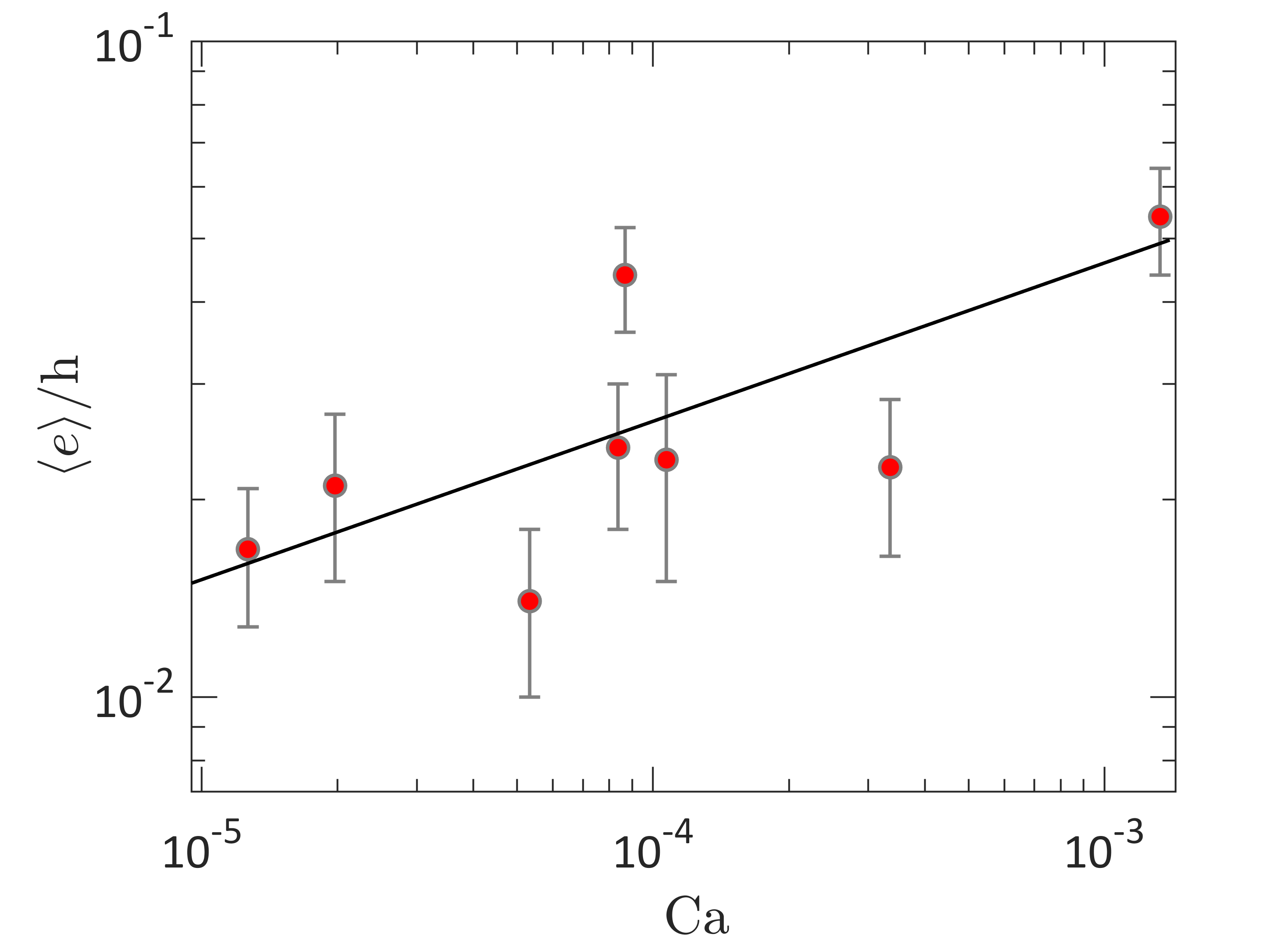}}
\caption{\small Non-dimensional film thickness between droplet and capillary boundary growth as a function of capillary number for $k=1.8\pm0.2$.}
\label{fig5}
\end{figure}

\subsection{Underlying mechanism of the emergence of interfacial waves}
To elucidate the mechanisms underlying these oscillatory dynamics, we systematically analyzed droplet behavior under the varied experimental conditions summarized in Table \ref{table1}. Under confinement, the 5CB droplet can be well approximated as a cylindrical liquid column. According to the classical Rayleigh–Plateau instability, a liquid column of radius $R$ (here 250 $\mu$m) becomes unstable to surface-tension-driven perturbations only when the wavelength $\lambda_c$ satisfies $\lambda_c > 2\pi R$ ($\sim$ 1550~$\mu$m). Such long-wavelength disturbances typically grow over time, leading to column breakup into spherical droplets to minimize interfacial energy \cite{barker2023fluctuating}. In contrast, the characteristic wavelength of the oscillations observed in our swimming 5CB droplets ($\lambda_c \sim 350$–$700~\mu\text{m}$) is significantly shorter than the axial circumference (see figure S4 in Supporting Information). This pronounced disparity indicates that the observed deformations are not due to Rayleigh–Plateau instability. 

Extensive research into active liquid crystals has identified interfacial instabilities, such as traveling waves, while highlighting the critical role of orientational order in driving these phenomena \cite{gulati2024traveling, sessa2025interfacial, soni2019stability, wu2017transition, adkins2022dynamics, zhao2024asymmetric}. In systems possessing a nematic base state, even marginal activity can amplify molecular bend fluctuations, triggering interfacial instabilities and wave propagation. As reported by \citet{gulati2024traveling}, local nematic order emerges when the activity-induced shear rate, $1/\tau \sim |\alpha_a|/\eta_{\mathrm{LC}}$, overcomes the intrinsic molecular relaxation rate $1/\tau_{\mathrm{LC}}$. Here, $\alpha_a \sim v \eta_{\mathrm{LC}}/w$ is the active stress while $\tau_{\mathrm{LC}}$ is governed by rotational viscosity $\nu \sim \eta_{\mathrm{LC}}$ \cite{gulati2024traveling, soni2019stability}, with $\eta_{\mathrm{LC}} = 17~\mathrm{mPa\,s}$ \cite{kimura2022probing}. For 5CB in nematic state, the molecular relaxation time scale can be estimated as $\tau_{\mathrm{LC}} = 8\pi \nu d^3 / k_B T$, where $T = 328~\mathrm{K}$, $k_B = 1.38 \times 10^{-23}~\mathrm{J\,K^{-1}}$, and $d \sim 2~\mathrm{nm}$ denotes the characteristic length of the rod-like 5CB molecules, yielding $\tau_{\mathrm{LC}} \sim 500~\mathrm{ns}$. The shear-induced time scale $\tau$, corresponding to the maximum velocity gradient ($\sim 160~\mu\mathrm{s}^{-1}$), is obtained as $3 ~\mathrm{s}$. Since $\tau_{\mathrm{LC}} \ll \tau$, any locally induced nematic ordering relaxes rapidly and cannot be sustained, indicating that it is not the nematic order that contributes to the emergence of traveling waves at the droplet interface in our experiments.

\begin{figure}[t]
\centering {\includegraphics[scale=0.7]{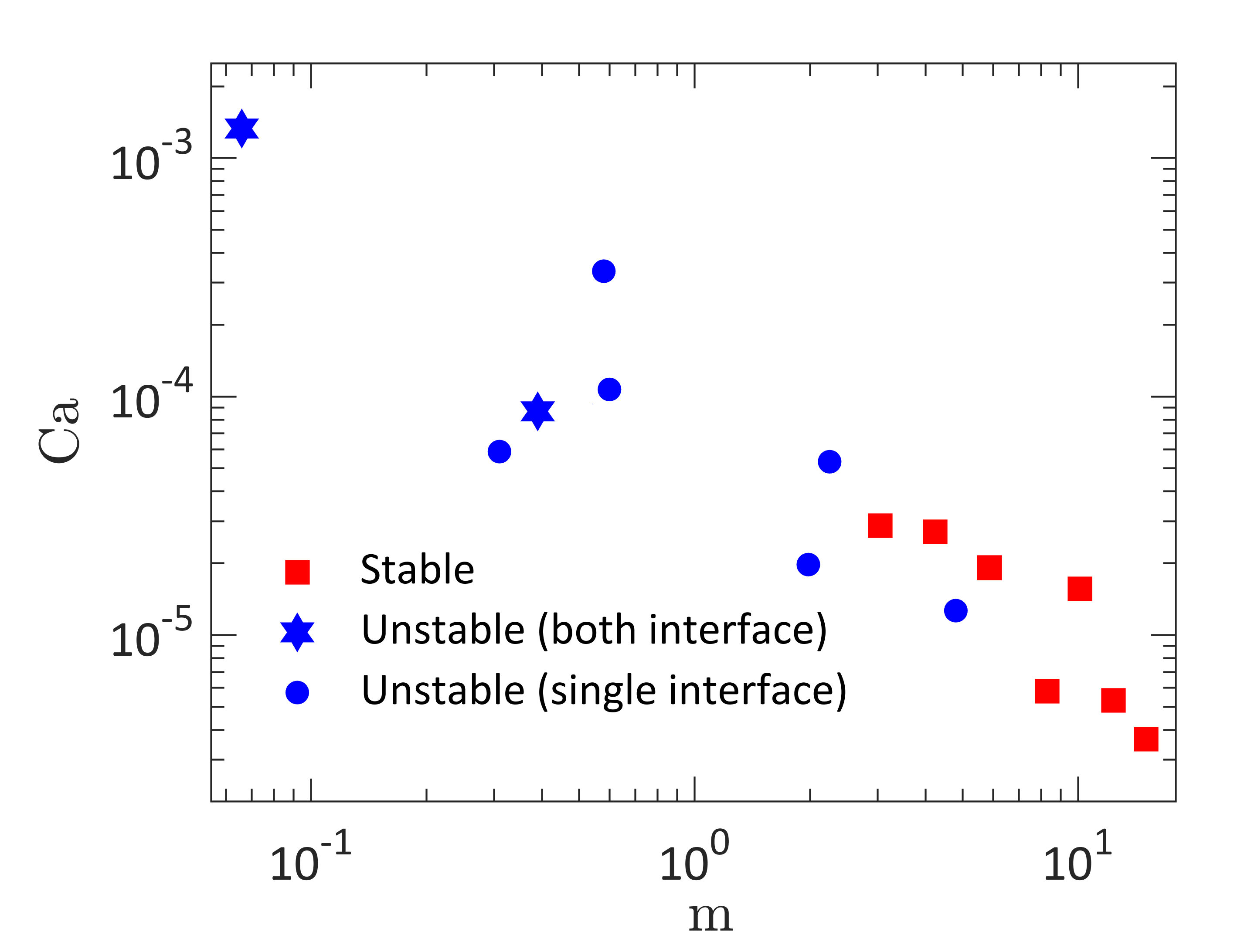}}
\caption{\small Phase space distribution of stable (steady interface) and unstable (traveling waves) between capillary number and viscosity ratio for $k=1.8\pm0.2$.}
\label{fig6}
\end{figure}

In the case of isotropic liquid crystals, interfacial instabilities arise only when the activity exceeds a critical threshold, beyond which local nematic order can be induced through flow alignment.  \citet{soni2019stability} demonstrated that, when the interface relaxes more slowly than the molecular phase, i.e., $\tau_s / \tau_{\mathrm{LC}} \gg 1$, instabilities and oscillations occur only for wavenumbers satisfying the scaling $k w \sim \tau_s / \tau_{\mathrm{LC}}$ \cite{soni2019stability}, where $\tau_s = \eta_{\mathrm{LC}} w / \sigma$ is the interfacial relaxation time. This scaling suggests that only oscillations with $\lambda_c$ $\ll$ 2$\pi d$ can develop in the LC film, where $d$ is film thickness. In contrast, in our experiments, the characteristic wavelength of the observed oscillations is of the order of the channel {width}, {i.e., 350--600 $\mu$m} (see figure~S4 in the Supporting Information), indicating a distinct instability mechanism in confined LC droplets.

\begin{figure}[ht]
 \centering {\includegraphics[scale=0.55]{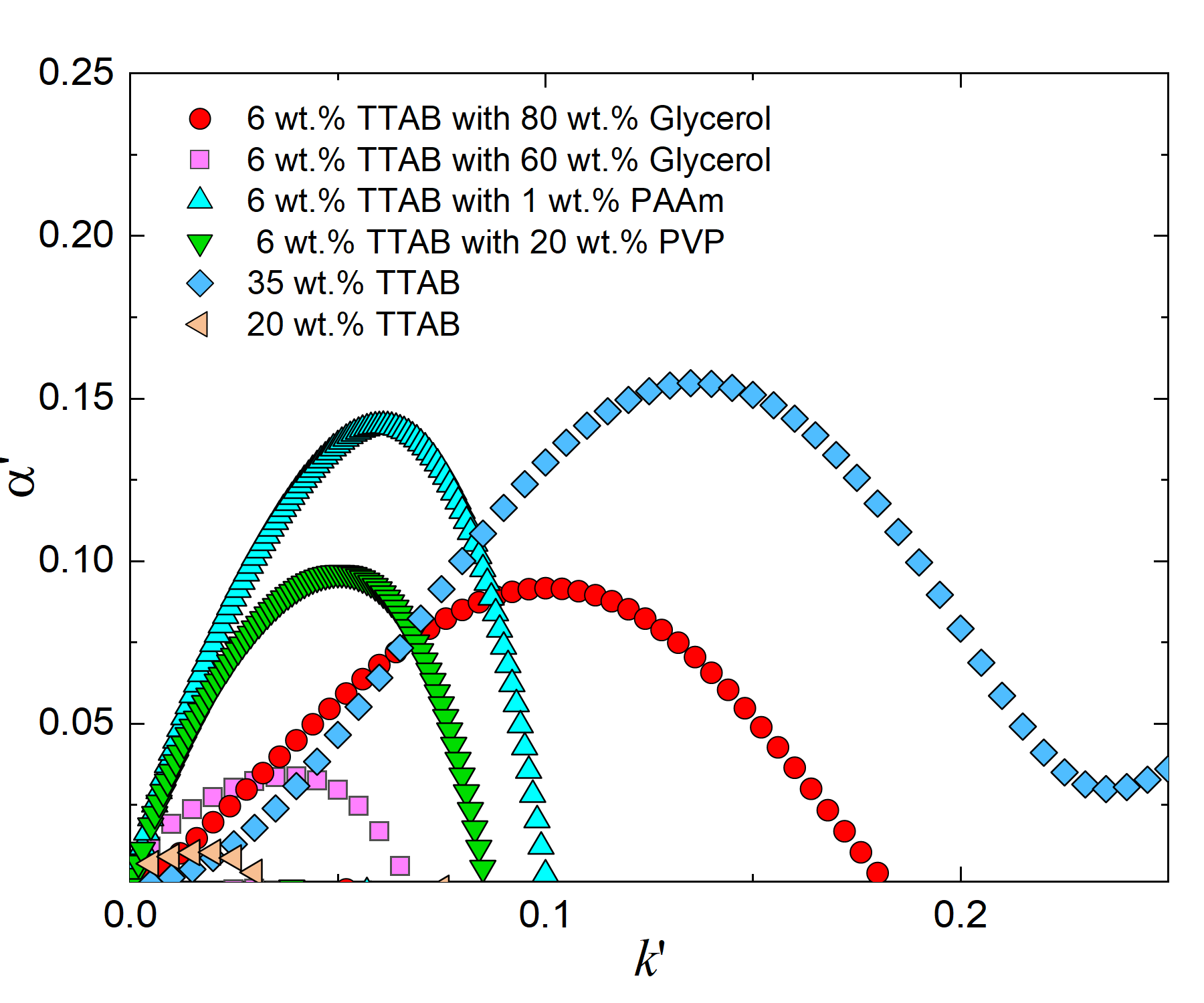}}
\caption{\small Non-dimensional growth rate versus non-dimensional wavenumber predicted for Yih--Marangoni instability \cite{picardo2016solutal} in cases where traveling waves are observed.}
\label{fig7}
\end{figure}

Another plausible mechanism underlying the observed interfacial traveling waves, is the classical Yih--Marangoni instability \cite{blyth2004effect}. In the absence of surfactants, stratified flow is known to become unstable to long-wave interfacial instability due to viscosity stratification. The instability first identified by \citet{yih1967instability}, occurs at any finite Reynolds number, howsoever small. The flow, however, is stable in the creeping flow limit, i.e., at zero Reynolds number \cite{yiantsios1988linear}. The long-wave asymptotic analysis reveals that the instability occurs only when the viscosity ratio, $m$, and the fluid layer thickness ratio, $n$ (both defined as the ratio of the droplet phase to the continuous phase), satisfy the criterion, $m < n^2$. In our experiments, Reynolds number is typically low with $\mathrm{Re} \sim 10^{-2}-0.1$. Further, the phase diagram in the capillary number–viscosity ratio parameter space (figure ~\ref{fig6}(B)) indicates that the droplet interface is increasingly susceptible to instability (blue markers) when the viscosity of the surrounding fluid is comparable to or greater than that of the droplet, i.e., $m>n^2$. Notably, this trend is observed at large capillary numbers. Furthermore, our linear stability analysis corresponding to the Yih mode carried out for the experimental conditions, discussed later, reveals that the range of unstable wavelengths are too long to manifest on the droplet interface. This suggests that the instability cannot be attributed to viscosity stratification alone. Instead, the presence of surfactants at the droplet interface is expected to play a significant role on the stability of the droplet interface. Various studies have reported that the presence of surfactants in immiscible two-layer channel flow can either stabilize or further destabilize the interface, depending on the system parameters \cite{halpern2003destabilization,gao2007effect,blyth2004effect,halpern2008nonlinear,frenkel2002stokes,kalogirou2020nonlinear,wei2005flow,jain2022instability,samanta2013effect,wei2005effects,kalogirou2019role,picardo2016solutal}. While most studies have been limited to insoluble surfactants, \citet{picardo2016solutal}, in particular, investigated the role of Marangoni flow generated by soluble surfactants on the Yih instability. \\ 
For the experimental conditions in which interfacial waves are observed, we perform a linear stability analysis of a stratified two-phase flow base state, incorporating surface-tension gradients, following the work of \citet{picardo2016solutal}. The results of the stability analysis reveal that the most unstable wavelength (i.e., the wavelength with the maximum growth rate) corresponds well with the experimentally observed values. This is evident from figure~\ref{fig7}, wherein the non-dimensional growth rate ($\alpha'$) versus non-dimensional wavenumber ($k'$), as predicted for Yih--Marangoni instability \cite{picardo2016solutal}, is plotted for the cases where traveling waves are observed. It is to be noted that the unstable case of 20wt\% TTAB lies just at the stability threshold, as is evident from the stability map in capillary number-viscosity ratio parameter space depicted in figure ~\ref{fig6}(B). Hence, except for this lone case, all other unstable cases that lie well within the unstable region are compared with the theory. The wavenumbers are non-dimensionalized using the corresponding lubricating film thickness, $\xi$, as the characteristic length scale. The dimensional wavelength for each case comes out to be of the order of a few hundred micrometers, which matches with the experimental observations, affirming that the Yih--Marangoni instability is a possible mechanism behind the onset of interfacial traveling waves.

\section{Conclusions}
This study elucidates how geometric confinement and interfacial physicochemical conditions regulate the motility and deformation dynamics of active 5CB droplets. In 6 wt.\% TTAB solution without and with introduction of solute, such as 80 wt.\% glycerol, droplet velocity exhibits a confinement-dependent decay, eventually saturating once a threshold is reached, which is expected to be attributed to the balance of hydrodynamic resistance and
Marangoni-driven slip. Despite the similar trend in propulsion velocity, the flow field distribution shows significant variations. Particularly, at the weaker confinement ratio $\sim 1$, the droplet propels in a consistent puller mode in 6 wt.$\%$, however, upon the introduction of 80 wt.\% glycerol, the swimming state transitions to temporal fluctuations in the flow field circulation at the anterior and posterior regions.

Notably, at elevated surfactant concentrations or in the presence of solutes such as glycerol, we observe a qualitatively distinct dynamical state: the emergence of traveling waves along the droplet interface. These bulging, peristaltic-like deformations propagate from the anterior to the posterior, a phenomenon absent in dilute surfactant systems. These waves manifest above a critical capillary number, where the gap thickness between the droplet and capillary boundary increases and the interfacial gradient becomes sufficient to destabilize the interface. Stability analysis of the two-layer liquid system confirms that the possible mechanism of traveling wave patterns on confined droplet is associated with Yih-Marangoni instability.

These findings demonstrate that confined droplets can achieve motility through traveling deformations, analogous to the euglenoid peristaltic motion observed in biological microswimmers~\cite{noselli2019swimming}. The synergy between confinement and Marangoni stresses thus identifies a novel physical mechanism for adaptive locomotion in constrained geometries. Beyond their biological implications, these results establish fundamental design principles for the development of synthetic soft microswimmers and active fluidic systems engineered for efficient transport in highly viscous or geometrically restricted environments.
\\

\section{acknowledgement}
We acknowledge the funding received by Department of Chemicals and Petrochemicals grant PC-II-25014/3/2022-PC II-CPC (FTS: 3019123), Science and Engineering Research Board CRG/2022/003763), Department of Science and Technology, India. We would like to thank Prof. Viswanathan Shankar, Department of Chemical Engineering, IIT Kanpur, for his constructive discussions.



\bibliography{arxiv}


\clearpage
\onecolumngrid          
\setcounter{figure}{0}

\renewcommand{\thefigure}{S\arabic{figure}}

\section*{Supporting Figures}
\begin{tabular}{c c}
\hline
\end{tabular}

\begin{figure}[h]
\centerline{\includegraphics[width=16cm]{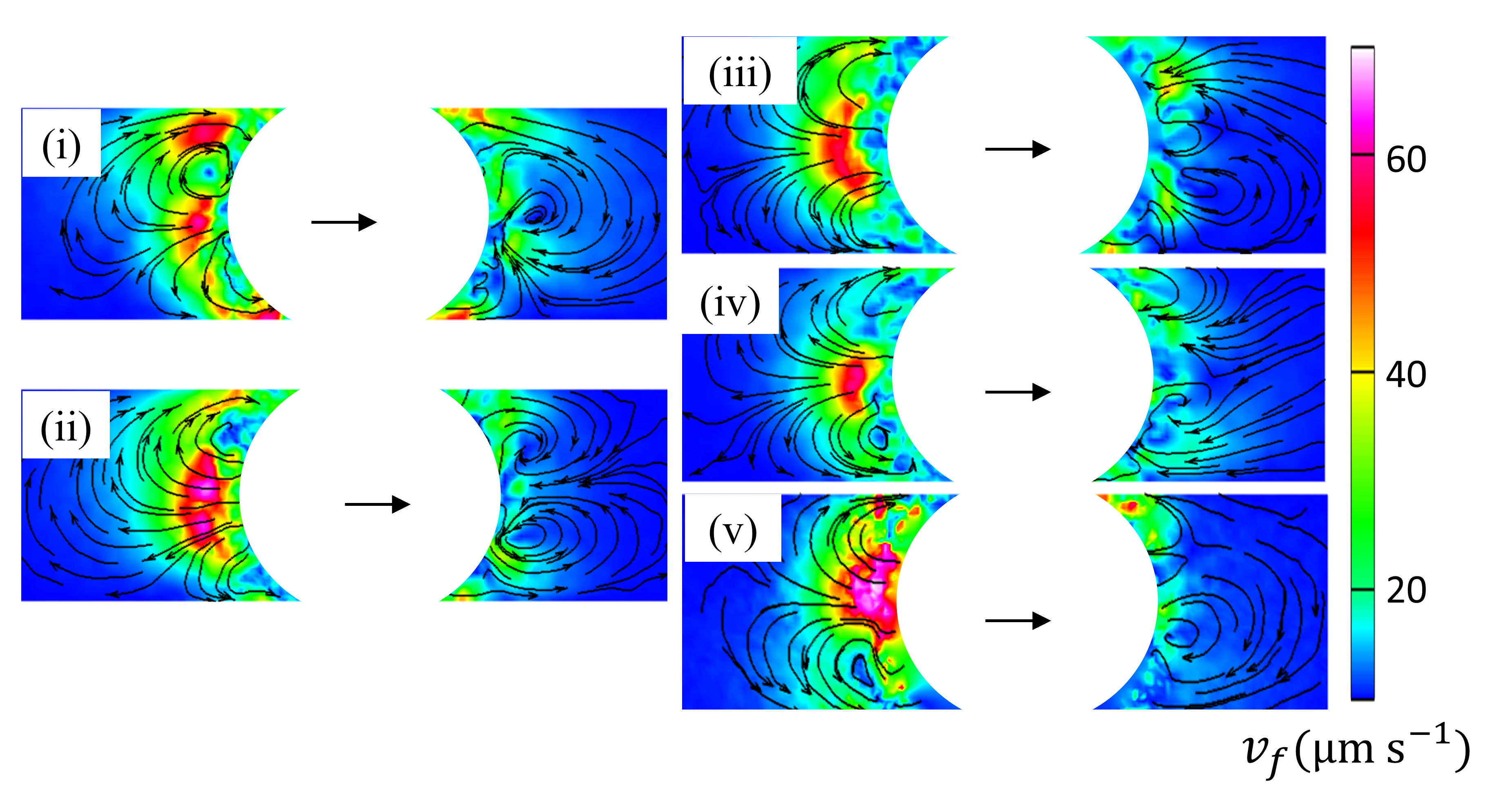}}
\linespread{1.0}
\caption{\small Flow velocity temporal fluctuations of 5CB droplet in 6 wt~$\%$ TTAB aqueous solution containing 80~$wt.\%$ glycerol for $k\sim1.1$. }
\label{1}
\end{figure}

\begin{figure}[h]
\centerline{\includegraphics[width=13cm]{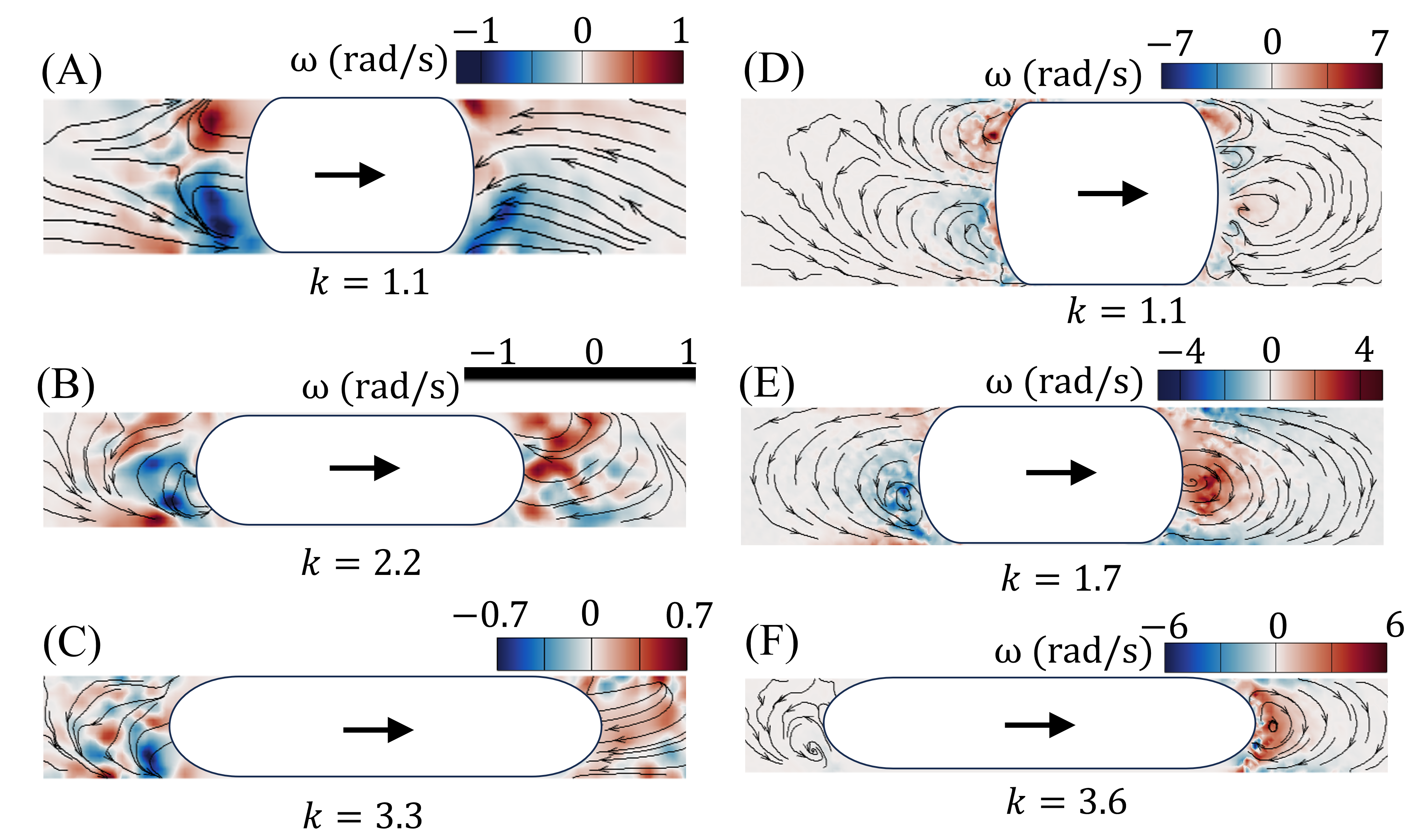}}
\linespread{1.0}
\caption{\small Experimentally determined vorticity ($\omega$) distribution, superimposed with flow field stream lines for the confined active droplet in 6 wt.$\%$ TTAB aqueous solution (A-C) without any additive and (D-F) with 80 wt.$\%$ glycerol for three different confinement ratios.}
\label{2}
\end{figure}

\begin{figure}[h]
\centerline{\includegraphics[width=10cm]{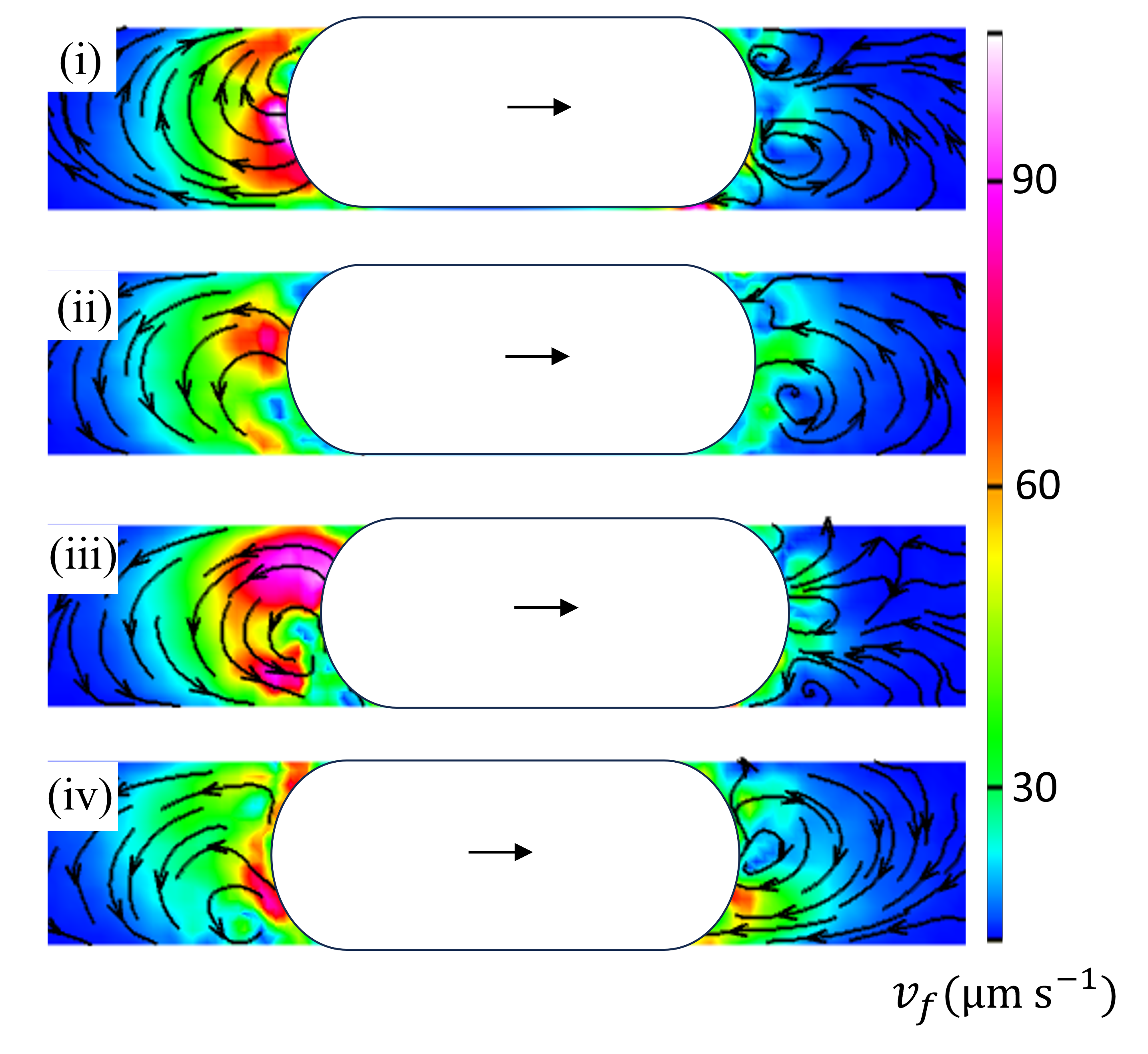}}
\linespread{1.0}
\caption{\small Flow velocity temporal fluctuations of 5CB droplet in 35 wt~$\%$ TTAB aqueous solution for $k\sim1.9$. }
\label{3}
\end{figure}

\begin{figure}[h]
\centerline{\includegraphics[width=10cm]{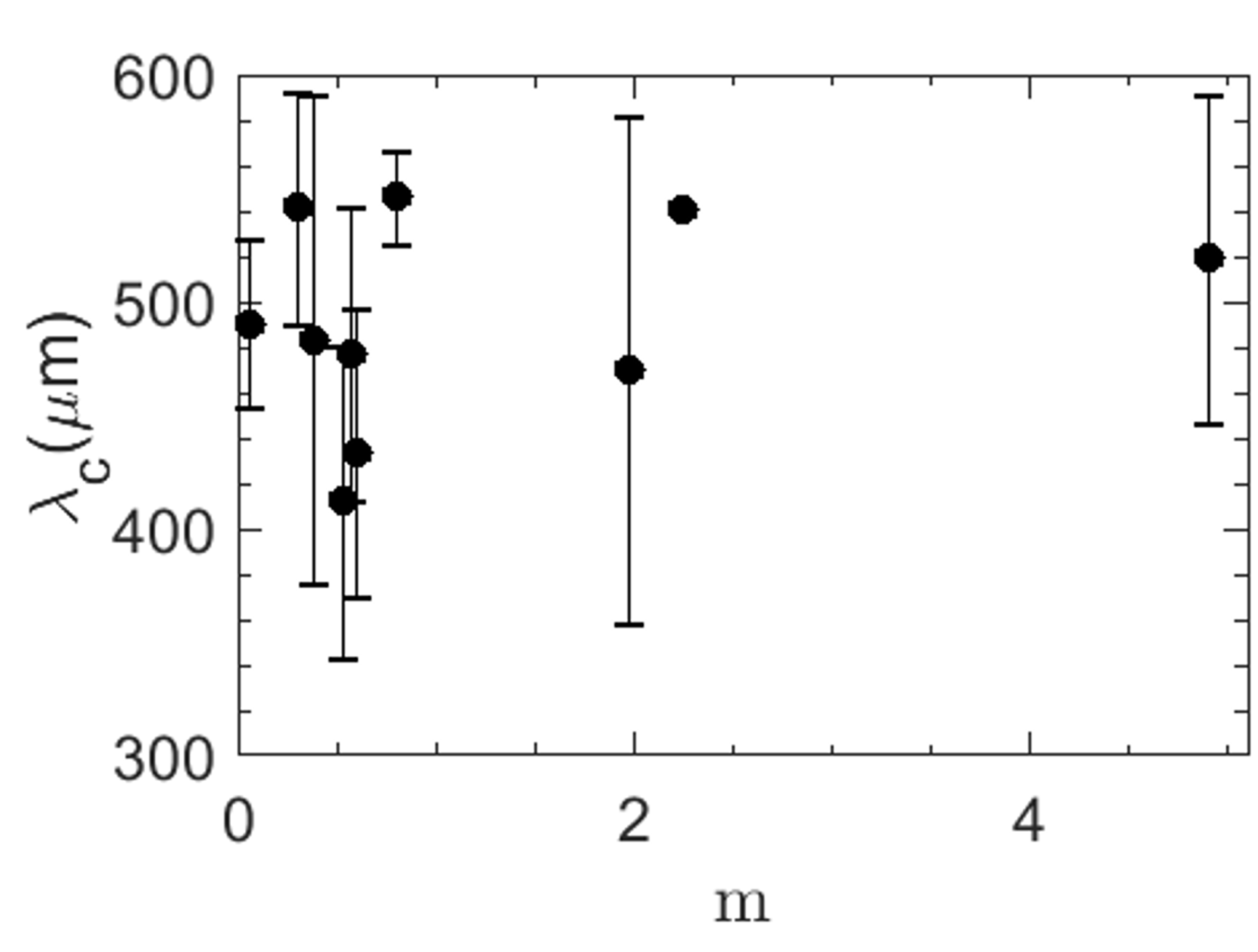}}
\linespread{1.0}
\caption{\small Average characteristic wavelength of droplet interface fluctuations for $k$ $\sim$ 1.9 as a function of viscosity ratio.}
\label{4}
\end{figure}




\end{document}